\documentclass[fleqn,usenatbib]{mnras}

\usepackage{newtxtext,newtxmath}

\usepackage[T1]{fontenc}
\usepackage{ae,aecompl}

\usepackage{graphicx}	
\usepackage{amsmath}	
\usepackage{amssymb}	
\usepackage{xspace}        

\newcommand{\target}{SDSS J090613.77$+$561015.2}

\title[Jet in a dwarf galaxy]{A parsec-scale radio jet launched by the central intermediate-mass black hole in the dwarf galaxy \target{}?}

\author[J. Yang et al.]{
Jun~Yang$^{1}$\thanks{E-mail: jun.yang@chalmers.se},
Leonid~I.~Gurvits$^{2,3,4}$,
Zsolt~Paragi$^{2}$,
S\'andor~Frey$^{5,6}$,
John~E.~Conway$^{1}$, 
\newauthor
Xiang~Liu$^{7}$ and
Lang~Cui$^{7}$
\\
\\
$^1$Department of Space, Earth and Environment, Chalmers University of Technology, Onsala Space Observatory, SE-43992 Onsala, Sweden\\
$^2$Joint Institute for VLBI ERIC, Oude Hoogevceensedijk 4, 7991 PD Dwingeloo, The Netherlands \\
$^3$Department of Astrodynamics and Space Missions, Delft University of Technology, Kluyverweg 1, 2629 HS Delft, The Netherlands \\
$^4$CSIRO Astronomy and Space Science, PO Box 76, Epping, NSW 1710, Australia \\
$^5$Konkoly Observatory, Research Centre for Astronomy and Earth Sciences, Konkoly Thege Mikl\'os \'ut 15-17, H-1121 Budapest, Hungary \\
$^6$Institute of Physics, ELTE E\"otv\"os Lor\'and University, P\'azm\'any P\'eter s\'et\'any 1/A, H-1117 Budapest, Hungary \\
$^7$Xinjiang Astronomical Observatory, Key Laboratory of Radio Astronomy, Chinese Academy of Sciences,150 Science 1-Street, 830011 Urumqi, P.R. China
}

\date{Accepted 2020 XXX. Received 2020 YYY; in original form 2020 ZZZ}

\pubyear{2020}

\begin{document}
\label{firstpage}
\pagerange{\pageref{firstpage}--\pageref{lastpage}}
\maketitle

\begin{abstract}
The population of intermediate-mass black holes (IMBHs) in nearby dwarf galaxies plays an important ``ground truth'' role in exploring black hole formation and growth in the early Universe. In the dwarf elliptical galaxy \target{} ($z=0.0465$), an accreting IMBH has been revealed by optical and X-ray observations. Aiming to search for possible radio core and jet associated with the IMBH, we carried out very long baseline interferometry (VLBI) observations with the European VLBI Network (EVN) at 1.66~GHz. Our imaging results show that there are two 1-mJy components with a separation of about 52~mas (projected distance 47~pc) and the more compact component is located within the 1$\sigma$ error circle of the optical centroid from available \textit{Gaia} astrometry. Based on their positions, elongated structures and relatively high brightness temperatures, as well as the absence of star-forming activity in the host galaxy, we argue that the radio morphology originates from the jet activity powered by the central IMBH. The existence of the large-scale jet implies that violent jet activity might occur in the early epochs of black hole growth and thus help to regulate the co-evolution of black holes and galaxies.
\end{abstract}





\section{Introduction}
\label{sec1}

The population of low-mass black holes (BHs) in nearby dwarf galaxies, i.e. galaxies with $M_\mathrm{g}\leq10^{9.5}M_{\sun}$, plays a key role in shedding light on BH formation and growth in the early Universe. One of the reasons of such the role is related to the fact that the probability of undergoing merging events in the population of dwarf galaxies is lower than for their larger counterparts, and therefore the masses of their central BHs are likely to remain close to their ``birth'' values \citep[e.g.][]{Reines2016, Greene2020}. These BHs with masses of $10^2 M_{\sun} \leq M_\mathrm{bh} \leq 10^6 M_{\sun}$ are typically classified as intermediate-mass black holes (IMBHs). Finding and weighing them would enable us to distinguish betwen different BH seed formation mechanisms, those involving a direct collapse at the mass level of  $M_\mathrm{bh}\sim10^4M_{\sun}$ and population III star seeds with a typical mass of $M_\mathrm{bh}\sim10^2M_{\sun}$ \citep[e.g., the simulation by][]{Volonteri2010}. 

At present, there are several hundred accreting IMBH candidates in dwarf galaxies that exhibit optical spectroscopic or X-ray signatures \citep[fraction $<1$ per cent,][]{Reines2013, Pardo2016}. Among them, only a small number of IMBHs \citep[e.g., twelve in][]{Schutte2019} have been identified in dwarf galaxies hosting active galactic nuclei (AGN).

About 0.3 per cent of dwarf galaxies have radio counterparts \citep{Reines2020} in the FIRST \citep[Faint Images of the Radio Sky at Twenty Centimeters,][]{Becker1995} survey. High-resolution very long baseline interferometry (VLBI) observations of these radio counterparts provide direct insight in their nature and mechanism of emission, involving non-thermal radio jet/outflow activity. Both jets and wide opening angle winds are major ingredients in the feedback mechanisms which are reflected in the co-evolution between BHs and galaxy bulges, i.e. the $M_{\rm bh}$--$\sigma$ correlation in which $\sigma$ is the stellar velocity dispersion \citep[e.g.][]{Greene2020, Kormendy2013} and the $M_{\rm bh}$--$M_{\rm bulge}$ correlation \citep[e.g.,][]{Schutte2019}. Revealing radio AGN activity would enable us to probe the feedback of the IMBHs \citep[e.g.][]{Greene2020, Manzano-King2019}. In addition, VLBI detections of compact radio cores would provide data points for filling the gap between supermassive and stellar-mass BHs for testing the mass-dependent relations \citep[e.g. the fundamental plane relation, ][]{Yuan2014} and allow us to search for off-nuclear IMBHs \citep{Reines2020} formed by galaxy mergers \citep[e.g.][]{Bellovary2019}. To date, there are some high-resolution imaging observations of IMBHs, e.g. POX 52 \citep{Thornton2008}, Henize~2--10 \citep{Reines2012} and NGC~404 \citep{Paragi2014}. However, radio jets or steady radio-emitting polar outflows, compact on sub-pc scales, have been revealed in only one dwarf galaxy, NGC~4395 \citep{Wrobel2006}.

The dwarf elliptical galaxy \target{} at the redshift $z=0.0465$ hosts an AGN and has a stellar mass of $2.3\times10^{9} M_{\sun}$ \citep[source ID: 9, ][]{Reines2013}. It displays not only some narrow-line AGN signatures but also a persistent broad H$\alpha$ emission \citep[source ID: RGG 9,][]{Baldassare2016}. Based on high spectral resolution observations, \citet{Baldassare2016} estimated the mass of its BH as $M_{\rm bh}=3.6^{+5.9}_{-2.3}\times10^5 M_{\sun}$ (including the systematic uncertainty of 0.42 dex). In the long-slit spectroscopy with the Keck I telescope, it shows some spatially extended ionized gas outflows that are most likely AGN-driven because their velocity (701$\pm$7 km~s$^{-1}$) exceeds the escape velocity (303$\pm$35 km~s$^{-1}$) of its halo \citep{Manzano-King2019}. It is a slightly resolved point-like source with total the flux density of 22.4$\pm$4.1~mJy in the the GMRT (Giant Metrewave Radio Telescope) 150~MHz all-sky radio survey \citep{Intema2017} and 4.7$\pm$0.2~mJy in the 1.4~GHz FIRST survey \citep{Becker1995}. It has a flux density of 0.93$\pm$0.05~mJy at 9.0~GHz and 0.78$\pm$0.04~mJy at 10.65~GHz in the Karl G. Jansky Very Large Array (VLA) observations \citep[source ID: 26,][]{Reines2020}.    

This Letter is composed as follows. In Section~\ref{sec2}, we describe our European VLBI Network (EVN) observations and data reduction. In Section~\ref{sec3}, we present the high-resolution imaging results. In Section~\ref{sec4}, we discuss the physical nature of the detected components and the implication from our findings. Throughout the paper, a standard $\Lambda$CDM cosmological model with H$_\mathrm{0}$~=~71~km~s$^{-1}$~Mpc$^{-1}$, $\Omega_\mathrm{m}$~=~0.27,  $\Omega_{\Lambda}$~=~0.73 is adopted; the VLBI images then have a scale of 0.9~pc mas$^{-1}$.

\begin{table}
\caption{Summary of the 1.66 GHz EVN observations of \target{}.  }
\label{tab1}
\begin{tabular}{cl}
\hline\hline
  Observing date and time       & Project code and participating stations       \\     
\hline     
2017 Jan 17, 18h--20h UT & RSY05: JbWbEfMcO8TrT6                        \\
2017 Nov 6, 00h--12h UT & EY029: JbWbEfMcO8TrT6UrSvZcBd   \\
\hline
\end{tabular}
\end{table}

\begin{figure}
\centering
\includegraphics[width=0.47\textwidth, clip=true]{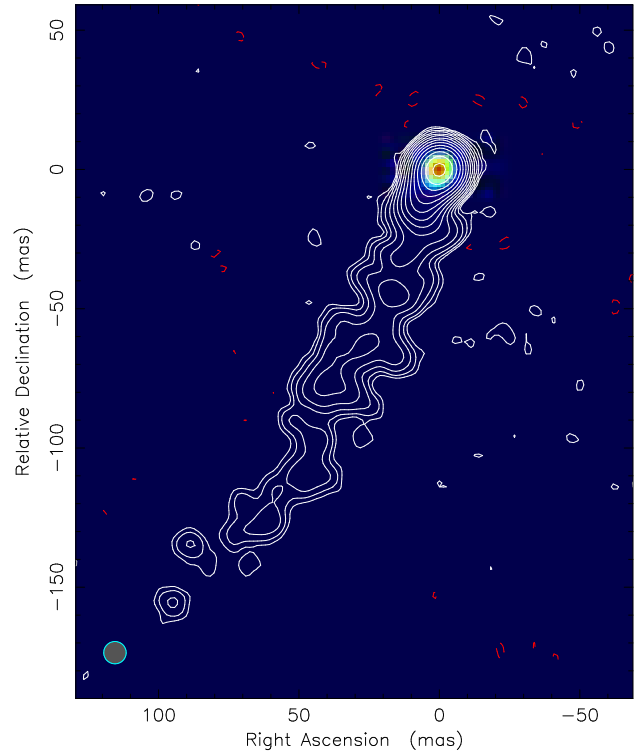}  \\
\caption{
The jet structure of the phase-referencing calibrator J0854$+$5757.  The 1.66-GHz EVN image has a dynamic range of $\frac{S_{\rm pk}}{\sigma_{\rm rms}}=$30\,250. The contours start at 0.016 mJy beam$^{-1} ($3$\sigma$) and increase by a factor of 2. The peak brightness is 484~mJy beam$^{-1}$. To properly show the faint jet structure, we used a large circular restoring beam with a FWHM of 8~mas.}
\label{fig1}
\end{figure}

\section{VLBI observations and data reduction}
\label{sec2}

We observed \target{} for two hours on 2017 January 17 with the e-EVN and for twelve hours on 2017 November 6 with the full EVN. Both observations used the standard 1-Gbps experiment setup (16 subbands, 16~MHz bandwidth per sub-band, dual circular polarisation, 2-bit quantisation) at the frequency of 1.66~GHz. The participating telescopes were Jodrell Bank Mk2 (Jb), Westerbork (Wb, single dish), Effelsberg (Ef), Medicina (Mc), Onsala (O8),  Tianma (T6), Urumqi (Ur), Toru\'n (Tr), Svetloe (Sv), Zelenchukskaya (Zc), Badary (Bd). Table~\ref{tab1} gives the observing time and the used stations at each epoch. The correlation was done by the EVN software correlator \citep[SFXC,][]{Keimpema2015} at JIVE (Joint Institute for VLBI, ERIC) using standard correlation parameters of continuum experiments.

Both experiments were performed in the phase-referencing mode to gain the calibration solutions and a reference position for our faint target. The bright source J0854$+$5757, about $2\fdg4$ apart from \target{} and a key source in the International Celestial Reference Frame \citep{Ma1998}, was observed periodically as the phase-referencing calibrator. The calibrator has a J2000 poisition at RA~$=08^{\rm h}54^{\rm m}41\fs996408$ ($\sigma_{\rm ra}=0.2$~mas), Dec.~$=57\degr57\arcmin29\farcs93914$ ($\sigma_{\rm dec}=0.1$~mas) in the source catalogue of 2015a from the Goddard Space Flight Centre VLBI group\footnote{ \url{http://astrogeo.org/vlbi/solutions/rfc_2015a/}}. The calibrator position has an offset of 0.7 mas with respect to the optical position in the second data release \citep[DR2,][]{Brown2018} of the \textit{Gaia} mission \citep{Prusti2016}. The nodding observations used a duty-cycle period of about five minutes (1~min for J0854$+$5757, 3~min for \target{}, 1~min for two scan gaps). During the observations, the sources was at an elevation of $\geq$18$\degr$ at all European telescopes.

The visibility data were calibrated using the National Radio Astronomy Observatory (NRAO) Astronomical Image Processing System \citep[\textsc{aips} version 31DEC17,][]{Greisen2003} software package. We removed the visibility data of side channels because of their low correlation amplitude while loading the data into \textsc{aips} and then ran the task \textsc{accor} to re-normalize the correlation amplitude. \textsc{A priori} amplitude calibration was performed with the system temperatures and the antenna gain curves if provided by the telescopes. If these data were missing, the nominal values were used. The ionospheric dispersive delays were corrected according to the map of total electron content provided by the Global Positioning System (GPS) satellite observations. Phase errors due to the antenna parallactic angle variations were removed. After a manual phase calibration was carried out, the global fringe-fitting and bandpass calibration were performed.

The calibrator J0854$+$5757 was imaged using iterations of model fitting with a group of point sources (delta functions) and self-calibration (Stokes \textit{I}) in the software package \textsc{difmap} \citep[version 2.5e, ][]{Shepherd1994}, fringe-fitting and self-calibration (Stokes \textit{RR} and \textit{LL}) in \textsc{aips}. The calibrator had a single-sided core--jet structure with a total flux density of 0.61$\pm$0.03~Jy in the first observing epoch. Due to a firmware bug in the European digital base-band converters, there were significant sensitivity losses in the 2nd epoch. According to the long-term light curve at 15 GHz observed by the 40-meter telescope at the Owens Valley Radio Observatory \citep{Richards2011} and published online\footnote{\url{http://www.astro.caltech.edu/ovroblazars}}, the calibrator had stable flux densities in 2017. Assuming that the calibrator had no flux density variation between the two epochs, we derived an amplitude correction factor of 1.63 in the imaging procedures. We used the jet base, the brightest component, as the reference point in the phase-referencing calibration. After about ten iterations, the deconvolved map of Stokes $I$ using natural weighting reached an image noise level of $0.016$~mJy~beam$^{-1}$, as low as the map of zero-flux-density Stokes $V$. The core--jet structure in the phase-referencing calibrator J0854$+$5757 observed in the second epoch is shown in Fig.~\ref{fig1}. In the final high dynamic range image, 82 positive point sources were used. Both the phase and amplitude self-calibration solutions were also transferred and applied to the target source. In the residual map of the target source, there are no clearly-seen systematic errors (noise peaks, strips and rings). This indicates that the phase-referencing calibration worked properly.

\begin{table*}
\caption{Summary of the circular Gaussian model-fitting results. Columns give (1) epoch, (2) component name, (3) peak brightness, (4) integrated flux density, (5--6) relative offsets in right ascension and declination with respect to component N, (7) deconvolved angular size (FWHM), (8) brightness temperature and (9) radio luminosity.}
\label{tab2}
\begin{tabular}{ccccrrccc}
\hline\hline
Epoch  &  Name   & $S_{\rm pk}$           & $S_{\rm obs}$        &  $\Delta\alpha\cos\delta$  &    $\Delta\delta$        & $\theta_{\rm size}$        &     $T_{\rm b}$   & $L_{\rm R}$       \\
            &               & (mJy beam$^{-1}$) &   (mJy)                    &  (mas)                               &      (mas)                    & (mas)                            &   (K)                           & erg s$^{-1}$     \\
\hline
1 &  N         &   0.616$\pm$0.022       & 0.94$\pm$0.06       & $0.00\pm$0.20                &  $0.00\pm$0.28           &   2.14$\pm$0.12          &     9.5$\times10^7$     & 7.7$\times10^{37}$ \\
1 &  S        &   0.428$\pm$0.022      & 1.04$\pm$0.06       & $-21.06\pm$0.44             &  $-51.14\pm$0.57       &   4.10$\pm$0.32          &     2.9$\times10^7$     & 8.5$\times10^{37}$ \\
\hline
2 &  N         &  0.781$\pm$0.016      & 0.88$\pm$0.02       & $0.00\pm$0.04                 &  $0.00\pm$0.05           &   1.82$\pm$0.14         &     1.2$\times10^8$   & 7.2$\times10^{37}$ \\
2 &  S        & 0.449$\pm$0.016      & 0.98$\pm$0.04       & $-22.47\pm$0.21             &  $-47.19\pm$0.21        &   8.02$\pm$0.42          &     7.1$\times10^6$  & 8.0$\times10^{37}$ \\
\hline
\end{tabular}  
\end{table*}

\section{Radio structure in \target{}}
\label{sec3}
\begin{figure}
\centering
\includegraphics[width=0.47\textwidth, clip=true]{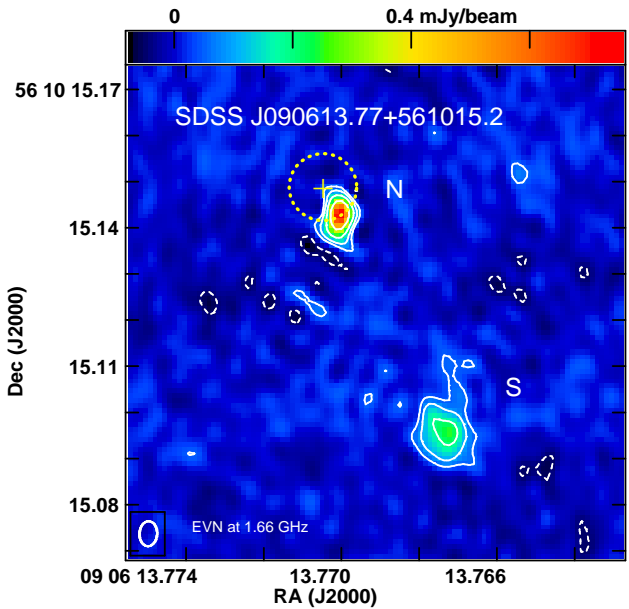}  \\
\caption{
A two-component brightness distribution found by the EVN at 1.66 GHz in the dwarf galaxy \target{} hosting an accreting IMBH. The yellow cross and circle mark the optical (\textit{Gaia} DR2) centroid and the total 1$\sigma$~error, respectively. The contours start at 0.048 mJy~beam$^{-1}$ (3$\sigma$) and increase by a factor of two. The peak brightness is 0.76 mJy~beam$^{-1}$. The restoring beam is 5.81~mas~$\times$~4.05~mas (FWHM) at $-3\fdg03$ position angle and plotted in the bottom-left corner.}
\label{fig2}
\end{figure}

The full-EVN image of \target{} obtained on 2017 November 6 is shown in Fig.~\ref{fig2}. There are two components detected and labelled as N and S in the \textsc{clean} map. The optical centroid, reported by the \textit{Gaia} DR2 \citep{Brown2018}, is marked as a yellow cross (J2000, RA $=09^{\rm h}06^{\rm m}13\fs77047$, Dec$=56\degr10\arcmin15\farcs1482$, $\sigma_{\rm ra}=\sigma_{\rm dec}=0.8$~mas, the astrometric excess noise to set the reduced $\chi^2_{\rm r}=1$ is 7.0~mas). The large excess noise is most likely related to the extended optical morphology and a certain level of asymmetric brightness distribution in the bulge \citep{Schutte2019}.  The total error, added in quadrature, is shown as a dotted yellow circle. The two components are also detected in the first short e-EVN observations. Due to its relatively low image quality, that image is not shown here. 

In order to determine the parameters of the obtained brightness distribution, we fitted two circular Gaussian components to the visibility data using \textsc{difmap}. The model-fitting results including the 1$\sigma$ formal uncertainties at the reduced $\chi_{\rm r}^2=1$ are listed in Table~\ref{tab2}. The systematic error of $S_{\rm pk}$, $S_{\rm obs}$ and $L_{\rm R}$ are about ten per cent. Using J0854$+$5757 as reference, we estimate the coordinates of the component N as RA$=09^{\rm h}06^{\rm m}13\fs77005$ and Dec$=56\degr10\arcmin15\farcs1429$ with a total systematic error $<1$ mas. The component N is partly resolved and has a faint ($\sim$0.2 mJy) extension of about 4~mas toward south. The separation between the components N and S is 52.3$\pm$0.3 mas. With respect to the component N, the component S has a position angle of about $-154\degr$. 

\begin{figure}
\centering
\includegraphics[width=0.47\textwidth, clip=true]{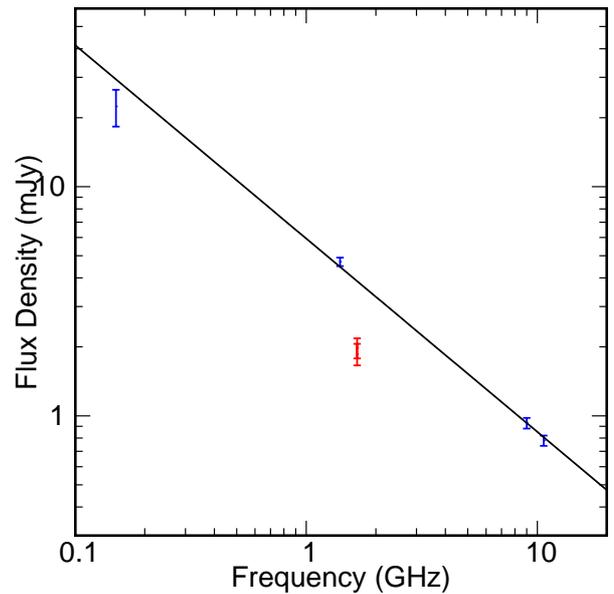}  \\
\caption{
The broad-band radio spectrum of \target{}. The blue points are from the earlier total flux density observations by the VLA \citep{Becker1995, Reines2020} and the GMRT \citep{Intema2017}. The two red points are from our high-resolution EVN observations. The black line shows the best-fit power-law spectrum to the low-resolution total flux density measurements (blue points).}
\label{fig3}
\end{figure}

All the available to date total flux density measurements of the source \target{} are plotted in Fig.~\ref{fig3}. Assuming no flux density variability, we fit the blue data points collected from the literature \citep{Becker1995, Intema2017, Reines2020} to a power-law spectrum $S_\nu=S_{0} \nu^\alpha$ and determine $S_0=5.94\pm0.39$ mJy, the spectral index $\alpha=-0.84\pm0.04$. According to this model, \target{} has a total flux density of 3.9$\pm$0.3 mJy at 1.66~GHz. Compared to this estimate, the high-resolution EVN image recovers only about 50 per cent. We also tried to search for diffuse radio emission. On the shortest baseline Ef--Wb, there is a hint for a faint and diffuse structure connecting the two components and extending farther on both sides. However, because of the lack of the shorter baselines, it is hard to make a reliable image for the diffuse structure from the available data. The existence of the diffuse radio structure is also expected since the source is slightly resolved (deconvovled FWHM: $2\farcs1 \times 1\farcs1$ in the major axis position angle 40$\degr$) in the VLA FIRST image \citep{Becker1995} and the elongation direction is roughly consistent with the overall EVN morphology extent. 

The next to last column in Table~\ref{tab2} presents an average brightness temperature, estimated as \citep[e.g.,][]{Condon1982},
\begin{equation}
T_{\rm b} = 1.22\times10^{9}\frac{S_\mathrm{obs}}{\nu_\mathrm{obs}^2\theta_\mathrm{size}^2}(1+z),
\label{eq1}
\end{equation}
where $S_\mathrm{obs}$ is the observed total flux density in mJy, $\nu_\mathrm{obs}$ is the observing frequency in GHz, $\theta_\mathrm{size}$ is the full width at half-maximum (FWHM) of the circular Gaussian model in mas, and $z$ is the redshift. The components N and S have average brightness temperatures of 1.2$\times10^8$ K and 7.1$\times10^6$ K, respectively, at 1.66~GHz in the second-epoch 12-h full-EVN observations. Due to the very limited $(u,v)$ coverage and the low image sensitivity, the component S has an underestimated $\theta_{\rm size}$ and thus an overestimated $T_{\rm b}$ in the first-epoch 2-hour-long e-EVN observation.

\section{Discussion}
\label{sec4}
\subsection{The nature of the components N and S}
\label{sec4-1}
The only plausible explanation of the radio structure in \target{} appears to be an AGN manifestation. The optical spectroscopic observations of \target{} show that there are no signatures for on-going star-forming activity in the BPT \citep{Baldwin1981} diagrams formed with some emission line ratios \citep{Reines2020}. The persistent broad H$\alpha$ emission is consistent with an AGN origin \citep{Baldassare2016}. Therefore, we reject a possibility that the observed components represent a superposition of supernova remnants (SNRs) like in, e.g., Arp 220 \citep{Varenius2019}. Moreover, they cannot be explained as two young radio supernovae because of their radio structures resolved on the pc scales although their radio luminosities ($L_{\rm R}\sim8\times10^{37}~$erg~s$^{-1}$) are in the luminosity range of young radio supernovae \citep[maximum: $L_{\rm R}\sim5\times10^{38}$~erg~s$^{-1}$,][]{Weiler2002}.

The component N is either the radio core, i.e. the jet base, or a newly-emerging jet component. Its proximity (within the 1$\sigma$ positional error) to the optical centroid of \target{} is consistent with this hypothesis. There also exists a jet-like faint extension toward South in the component N when the image resolution in North-South is slightly improved to 4 mas. We can also fit the component N to two point sources with flux densities 0.73$\pm$0.02 and 0.19$\pm$0.02 mJy and a separation of 3.9$\pm$0.4 mas. 

The component S is most likely an expanding ejecta. Because of its large distance ($\sim$58.5~mas) to the \textit{Gaia} centroid, it cannot be explained as the radio core. Compared to the component N, the component S has the more extended structure and the lower brightness temperature. Moreover, its position and elongation (position angle about 36$\degr$) are roughly aligned with the extension of the component N.

The \textit{Gaia} positioning of the optical centroid close to the radio component N provides a strong indication on the nature of this component as the compact radio core. However, we cannot rule out that \target{} is a young compact symmetric object \citep[CSO, ][]{Wilkinson1994}. In this scenario, the radio components N and S are a pair of moving-out radio ejectae (or mini-lobes) within its host galaxy, and the radio core is located somewhere in–between and undetected. The latter would be still consistent with the \text{Gaia} position within its $3\sigma$ error. Assuming a typical separation speed of $0.2c$ among CSOs \citep[e.g.][]{An2012}, the pair of components would have a kinematic lifetime of $\sim$7.5$\times$10$^2$ yr. This is a typical value in young extragalactic radio sources \citep[e.g.][]{An2012b}. Steep spectra at $\ga$1~GHz are not unknown among CSOs, e.g. J0132$+$5620 and J2203$+$1007 \citep{An2012}. Another example is the CSO PKS~1117$+$146, which has a spectral index of ~$-$0.7 \citep{Torniainen2007} and a large angular separation between the opposite jet components,$\sim$70~mas \citep{Bondi1998}, and thus is similar to \target{}. A conclusive test on possible CSO identification of the source will be provided by its future multi-frequency and multi-epoch VLBI studies.

\subsection{Implications of the presence of a radio jet associated with the IMBH}
\label{sec4-2}
If the component N is the stationary radio core associated with the IMBH, \target{} will have a relatively high radio luminosity. The source has an X-ray luminosity of $L_\mathrm{X}=4.5\times10^{40}$ erg\,s$^{-1}$ in the 2--7~keV band \citep{Baldassare2017}. This leads to a ratio of $\frac{L_\mathrm{R}}{L_\mathrm{X}}=1.8\times10^{-2}$, much higher than the typical value $\sim10^{-5}$ in the radio-quiet Palomar--Green quasar sample \citep{Laor2008}. Due to the high ratio, it is hard to explain them as wide-angle radio-emitting winds. At the low accretion rate state, there exists a correlation \citep[e.g.][]{Merloni2003} between the radio core luminosity at 5 GHz, the X-ray luminosity ($L_{\rm X}$) in the 2--10~keV band, and the BH mass ($M_{\rm bh}$):
\begin{equation}
\log L_{\rm R} =(0.60^{+0.11}_{-0.11}) \log L_{\rm X} + (0.78^{+0.11}_{-0.09}) \log M_{\rm bh} + 7.33^{+4.05}_{-4.07} 
\label{eq2}
\end{equation}
According to Equation~\ref{eq2}, we would expect $L_{\rm R}=10^{36.1\pm0.9}$ erg\,s$^{-1}$. This estimate would be one order of magnitude lower while still in the acceptable range in view of the large scatter of the correlation. 

The radio jet is rarely-seen in dwarf galaxies. Compared to the first and only case, NGC~4395 \citep{Wrobel2006}, \target{} has a jet 160 times longer and 10$^5$ times more luminous. The finding of the large jet structure indicates that violent ejections might appear at the BH growth stage of $M_{\rm bh}\sim$10$^5M_{\sun}$ in the early Universe. Most of these jets associated with low-mass BHs might be short-lived and sub-kpc objects because they have rather low radio luminosities \citep[$L_{\rm R}\leq10^{41}$ erg~s$^{-1}$ at 1.4 GHz,][]{Kunert2010, An2012b}.    

\citet{Manzano-King2019} report some AGN-driven outflows in dwarf galaxies, indicating significant AGN impact on the large-scale kinematics and gas content. The kpc-scale high-velocity ionized gas outflows in \target{} might be driven by the AGN jet activity, in particular the diffuse structure completely resolved out in our EVN image because of missing the shorter baselines. The unseen kpc-scale (relic) jet component has a flux density of 2.0$\pm$0.2 mJy. According the scaling relation $P_{\rm jet} = 5.8 \times10^{43} (\frac{L_{\rm R}}{10^{40}})^{0.70}$ erg\,s$^{-1}$ between jet kinematic power and the radio luminosity derived by \citet{Cavagnolo2010} using \textit{Chandra} X-ray and VLA radio data of radio galaxies, the unseen relic jet has a power of $P_{\rm jet}=10^{42.6\pm0.7}$ erg\,s$^{-1}$, reaching about ten percent of the Eddington luminosity $L_{\rm Edd}=10^{43.6\pm0.4}$ erg\,s$^{-1}$ of the IMBH. Thus, the AGN jet activity might have significant impact on the host galaxy.

\section*{Acknowledgements}
\label{ack}
The European VLBI Network is a joint facility of independent European, African, Asian, and North American radio astronomy institutes. Scientific results from data presented in this publication are derived from the following EVN project codes: RSY05 and EY029.
This work has made use of data from the European Space Agency (ESA) mission {\it Gaia} (\url{https://www.cosmos.esa.int/gaia}), processed by the {\it Gaia} Data Processing and Analysis Consortium (DPAC, \url{https://www.cosmos.esa.int/web/gaia/dpac/consortium}). Funding for the DPAC has been provided by national institutions, in particular the institutions participating in the {\it Gaia} Multilateral Agreement. 
LIG acknowledges support by the CSIRO Distinguished Visitor Programme.
We thank the staff of the GMRT that made these observations possible. The GMRT is run by the National Centre for Radio Astrophysics of the Tata Institute of Fundamental Research.
This research has made use of data from the OVRO 40-m monitoring program \citep{Richards2011} which is supported in part by NASA grants NNX08AW31G, NNX11A043G, and NNX14AQ89G and NSF grants AST-0808050 and AST-1109911.





\begin{thebibliography}{99}
\bibitem[\protect\citeauthoryear{An et al.}{2012}]{An2012}
An T., et al., 2012, \apjs, 198, 5
\bibitem[\protect\citeauthoryear{An \& Baan}{2012}]{An2012b}
An T., Baan W.~A., 2012, \apj, 760, 77
\bibitem[\protect\citeauthoryear{Baldassare et al.}{2016}]{Baldassare2016}
Baldassare V.~F., et al., 2016, \apj, 829, 57
\bibitem[\protect\citeauthoryear{Baldassare et al.}{2017}]{Baldassare2017}
Baldassare V.~F., Reines A.~E., Gallo E., Greene J.~E., 2017, \apj, 836, 20
\bibitem[\protect\citeauthoryear{Baldwin, Phillips \& Terlevich}{1981}]{Baldwin1981}
Baldwin J.~A., Phillips M.~M., Terlevich R., 1981, \pasp, 93, 5
\bibitem[\protect\citeauthoryear{Becker, White \& Helfand}{1995}]{Becker1995}
Becker R.~H., White R.~L., Helfand D.~J., 1995, \apj, 450, 559
\bibitem[\protect\citeauthoryear{Bellovary et al.}{2019}]{Bellovary2019}
Bellovary J.~M., Cleary C.~E., Munshi F., Tremmel M., Christensen C.~R., Brooks A., Quinn T.~R., 2019, MNRAS, 482, 2913
\bibitem[\protect\citeauthoryear{Bondi et al.}{1998}]{Bondi1998}
Bondi M., Garrett M.~A., Gurvits L.~I., 1998, \mnras, 297, 559
\bibitem[\protect\citeauthoryear{Brown et al.}{2018}]{Brown2018}
Brown A.~G.~A., et al. (Gaia Collaboration), 2018, \aap, 616, A1
\bibitem[\protect\citeauthoryear{Cavagnolo et al.}{2010}]{Cavagnolo2010}
Cavagnolo K.~W., McNamara B.~R., Nulsen P.~E.~J., Carilli C.~L., Jones C, B\^irzan L, 2010, \apj, 720, 1066
\bibitem[\protect\citeauthoryear{Condon et al.}{1982}]{Condon1982}
Condon J.~J., Condon M.~A., Gisler G., Puschell J.~J., 1982, \apj, 252, 102
\bibitem[\protect\citeauthoryear{Greene, Strader \& Ho}{2020}]{Greene2020}
Greene J.~E., Strader J., Ho L.~C., 2020, \araa, in press, arXiv:1911.09678
\bibitem[\protect\citeauthoryear{Greisen}{2003}]{Greisen2003} 
Greisen E.~W., 2003, in Heck A., ed., Astrophysics and Space Science Library, Vol. 285, Information Handling in Astronomy: Historical Vistas. Kluwer, Dordrecht, p. 109
\bibitem[\protect\citeauthoryear{Intema et al.}{2017}]{Intema2017}
Intema H.~T., Jagannathan, P., Mooley~K.~P., Frail D.~A., 2017, \aap, 598, A78
\bibitem[\protect\citeauthoryear{Keimpema et al.}{2015}]{Keimpema2015}
Keimpema A., et al., 2015, Exp. Astron., 39, 259
\bibitem[\protect\citeauthoryear{Kormendy \& Ho}{2013}]{Kormendy2013}
Kormendy J., Ho L.~C., 2013, \araa, 143, 511 
\bibitem[\protect\citeauthoryear{Kunert-Bajraszewska et al.}{2010}]{Kunert2010}
Kunert-Bajraszewska M., Gawro'nski M.~P., Labiano A., Siemiginowska A., 2010, \mnras, 408, 2261
\bibitem[\protect\citeauthoryear{Laor \& Behar}{2008}]{Laor2008}
Laor A., Behar E.,  2008, \mnras, 390, 847
\bibitem[\protect\citeauthoryear{Ma et al.}{1998}]{Ma1998}
Ma C., Arias E.~F., Eubanks T.~M., Fey A.~L., Gontier A.-M., Jacobs C.~S., Sovers O.~J., Archinal B.~A., Charlot P., 1998, \aj, 116, 516
\bibitem[\protect\citeauthoryear{Manzano-King, Canalizo \& Sales}{2019}]{Manzano-King2019}
Manzano-King C., Canalizo G., Sales L.~V.,  2019, \apj, 884, 54
\bibitem[\protect\citeauthoryear{Merloni, Heinz \& Di Matteo}{2003}]{Merloni2003}
Merloni A., Heinz S., Di Matteo~T., 2003, \mnras, 345, 1057
\bibitem[\protect\citeauthoryear{Paragi et al.}{2014}]{Paragi2014} 
Paragi Z., Frey S., Kaaret P., Cseh D., Overzier R., Kharb P., 2014, \apj, 791, 2
\bibitem[\protect\citeauthoryear{Pardo et al.}{2016}]{Pardo2016} 
Pardo K., et al., 2016, \apj, 831, 203
\bibitem[\protect\citeauthoryear{Prusti et al.}{2016}]{Prusti2016}
Prusti T., et al. (Gaia Collaboration), 2016. \aap, 595, A1
\bibitem[\protect\citeauthoryear{Reines \& Comastri}{2016}]{Reines2016} 
Reines A.~E., Comastri A., 2016, \pasa, 33, e054
\bibitem[\protect\citeauthoryear{Reines et al.}{2020}]{Reines2020} 
Reines A.~E., Condon J.~J., Darling J., Greene J.~E., 2020, \apj, 888, 36
\bibitem[\protect\citeauthoryear{Reines \& Deller}{2012}]{Reines2012} 
Reines A.~E., Deller A.~T., 2012, \apj, 750, L24
\bibitem[\protect\citeauthoryear{Reines, Greene \& Geha}{2013}]{Reines2013}
Reines A.~E., Greene J.~E., Geha M., 2013, \apj, 775, 116
\bibitem[\protect\citeauthoryear{Richards et al.}{2011}]{Richards2011}
Richards J.~L., et al., 2011, \apjs, 194, 29
\bibitem[\protect\citeauthoryear{Schutte, Reines \& Greene}{2019}]{Schutte2019} 
Schutte Z., Reines A.~E., Greene J.~E., 2019, \apj, 887, 245
\bibitem[\protect\citeauthoryear{Shepherd, Pearson \& Taylor}{1994}]{Shepherd1994}
Shepherd M.~C., Pearson T.~J., Taylor G.~B., 1994, \baas, 26, 987
\bibitem[\protect\citeauthoryear{Thornton et al.}{2008}]{Thornton2008} 
Thornton C.~E., Barth A.~J., Ho L.~C., Rutledge R.~E., Greene J.~E., 2008, ApJ, 686, 892
\bibitem[\protect\citeauthoryear{Torniainen et al.}{2007}]{Torniainen2007}
Torniainen I., Tornikoski M., L{\"a}hteenm{\"a}ki A., Aller M.~F., Aller H.~D., Mingaliev M.~G., 2007, \aap, 469, 451
\bibitem[\protect\citeauthoryear{Varenius et al.}{2019}]{Varenius2019}
Varenius E., et al., 2019, \aap, 623, A173
\bibitem[\protect\citeauthoryear{Volonteri}{2010}]{Volonteri2010} 
Volonteri M., 2010, \araa, 18, 279
\bibitem[\protect\citeauthoryear{Weiler et al.}{2002}]{Weiler2002}
Weiler K.~W., Panagia N., Montes M.~J., Sramek R.~A., 2002, \araa, 40, 387
\bibitem[\protect\citeauthoryear{Wilkinson et al.}{1994}]{Wilkinson1994}
Wilkinson P.~N., Polatidis A.~G., Readhead A.~C.~S., Xu~W., Pearson T.~J., 1994, \mnras, 269, 67 
\bibitem[\protect\citeauthoryear{Wrobel \& Ho}{2006}]{Wrobel2006} 
Wrobel J.~M., Ho L.~C., 2006, \apj, 646, L95
\bibitem[\protect\citeauthoryear{Yuan \& Narayan}{2014}]{Yuan2014}
Yuan F., Narayan R., 2014, \araa, 52, 529 

\end{thebibliography}

%







\bsp	
\label{lastpage}
\end{document}